# Observation of flat-band localizations and topological edge states induced by effective strong interactions in electrical circuit networks

Xiaoqi Zhou[1*], Weixuan Zhang[1*,+], Houjun Sun[2], and Xiangdong Zhang[1$]

[1]Key Laboratory of advanced optoelectronic quantum architecture and measurements of Ministry of Education, Beijing Key Laboratory of Nanophotonics & Ultrafine Optoelectronic Systems, School of Physics, Beijing Institute of Technology, 100081, Beijing, China

[2] Beijing Key Laboratory of Millimeter wave and Terahertz Techniques, School of Information and Electronics, Beijing Institute of Technology, Beijing 100081, China

*These authors contributed equally to this work.

[+$]Author to whom any correspondence should be addressed. E-mail: zhangwx@bit.edu.cn
zhangxd@bit.edu.cn

**Flat-band topologies and localizations in non-interacting systems are extensively studied in different quantum and classical-wave systems. Recently, the exploration on the novel physics of flat-band localizations and topologies in interacting systems has aroused great interest. In particular, it is theoretically shown that the strong-interaction could drive the formation of nontrivial topological flat bands, even dispersive trivial bands dominate the single-particle counterparts. However, the experimental observation of those interesting phenomena is still lacking. Here, we experimentally simulate the interaction-induced flat-band localizations and topological edge states in electrical circuit networks. We directly map the eigenstates of two correlated bosons in one-dimensional Aharonov–Bohm cages to modes of two-dimensional circuit lattices. In this case, by tuning the effective interaction strength through circuits' groundings, the two-boson flat-bands and topological edge states are detected by measuring frequency-dependent impedance responses and voltage dynamics in the time domain. Our finding suggests a flexible platform to simulate the interaction-induced flat-band topology, and may possess potential applications in designing novel electronic devices.**

Exotic physics of topological flat bands, which allow the coexistence of strong spatial localizations and non-trivial topological boundary states, has gained much attentions in the past few decades. To date, a large number of theoretical models have been put forward to construct non-trivial topological flat bands. For example, topological flat bands with non-zero Chern numbers have been proposed by introducing the long-range couplings in original Haldane model [1, 2]. Lately, flat bands with arbitrary large Chern numbers were systematically created in multilayer systems with only short-range hopping [3, 4]. Except for the Chern-class topological flat bands, low-energy moiré flat bands with non-trivial valley Chern numbers have been generated in twisted bilayer structures [5-9]. Furthermore, the complete flat-band systems sustaining the square-root topological boundary states can be induced by flux-controlled destructive interferences in Aharonov–Bohm (AB) cages [10-17], which are wildly constructed in bipartite lattices, rhombus chains, and Creutz ladders. These interesting phenomena promote the experimental implementation of topological flat bands in both non-interacting quantum and classical systems [18-29].

Motivated by novel properties of the flat-band topology in single-particle systems, it is very interesting to explore the fate of topological flat bands in interacting systems. So far, the impact of particle interactions on properties of topological flat bands in AB-cages has been investigated at different regimes [30-38]. At the two-body level, it is shown that the interaction can induce the delocalization and spreading in complete flat-band systems. In particular, the property of spatial localizations was shown to be substantially altered by interactions, which can break the destructive interference of single particle states, and introduce a mechanism by which the two-particle wave function can spread over the entire lattice [30, 31]. At the many-body level, the Hubbard-type interaction can lift the degeneracy of flat bands, and produce a rich filling-factor-dependent quantum phases, including the super-solidity at low filling [32] and a pair condensation at commensurate filling [33, 34]. On the other hand, some theoretical studies have also shown that the strong two-body interaction can trigger two bosons to form doublons, which can themselves experience AB-caging at different values of the magnetic flux [35]. Although much theoretical investigations have been put forward, the suitable experimental platform to implement AB cages with interactions is still lacking. Hence, how to observe the interaction-induced flat-band localizations and topological edge states is still to be explored.

In this work, we experimentally emulate the interaction-induced two-boson flat-band localizations and topological edge states in circuit networks. The eigenstates of two strongly interacted bosons in one-dimensional AB cages are directly mapped to modes of two-dimensional circuit lattices, where the effective interaction strength can be tuned by the grounding of electric circuits. Based on our designed circuit simulators, the analogies of two-boson flat-bands and topological edge states have been observed by measuring the frequency-dependent impedance responses and the voltage dynamics in time domain. Our work suggests a flexible platform to simulate the few-body AB cage with interactions and may possess potential applications in novel electronic signal control.

**The theory of simulating interaction-induced flat-band localizations and topological edge states by circuit networks.** We consider a pair of correlated bosons hopping on a one-dimensional (1D) lattice with $N$ unit cells. The location of each lattice site is labeled by two indicators $(l, \alpha)$, where $\alpha (= A, B, C)$ mark three sublattices in each unit, and $l$ corresponds to the coordinate of the unit cell. Additionally, a synthetic magnetic flux is applied in each closed plaquette via Peierls' substitution of the coupling strength $(Je^{i\theta})$ between sublattice sites $A$ and $B$, as shown in Fig. 1a. In this case, the system can be described by the extended version of the 1D Bose-Hubbard Hamiltonian as:

$$H = -J \sum_{l=1}^{L} \left( e^{i\theta} a_{l,A}^+ a_{l,B} + a_{l,A}^+ a_{l,C} + a_{l,B}^+ a_{l+1,A} + a_{l,C}^+ a_{l+1,A} + h.c. \right) + 0.5U \sum_{l,\alpha} n_l(n_l - 1) \quad (1)$$

where $a_{l,\alpha}^+ (a_{l,\alpha})$ and $n_l = a_{l,\alpha}^+ a_{l,\alpha}$ ($\alpha = A, B, C$) are the bosonic creation (annihilation) and particle number operators, respectively. $J$ is the single-particle hopping rate between adjacent sites, and $U$ is the on-site interaction energy, as illustrated in Fig. 1b. The two-boson solution can be expanded in the Fock space as $|\psi\rangle = \frac{1}{\sqrt{2}} \sum_{m,n=1;\alpha,\beta}^{L} \varphi_{(m,\alpha)(n,\beta)} a_{(m,\alpha)}^+ a_{(n,\beta)}^+ |0\rangle$, where $\varphi_{(m,\alpha)(n,\beta)}$ is the probability amplitude to find one boson at site $(m, \alpha)$ and the other at site $(n, \beta)$. Solving the steady state Schrodinger equation $H|\psi\rangle = \varepsilon|\psi\rangle$, the eigen-equations with respect to $\varphi_{(m,\alpha)(n,\beta)}$ can be obtained. Eq. (2) corresponds to the eigen-equation with two bosons both locating at A sublattices

$$\varepsilon \varphi_{(m,A)(n,A)} = -J[e^{i\theta} \varphi_{(m,A)(n,B)} + \varphi_{(m,A)(n,C)} + \varphi_{(m,A)(n-1,C)} + \varphi_{(m,A)(n-1,B)}$$

$$+ e^{i\theta} \varphi_{(m,B)(n,A)} + \varphi_{(m,C)(n,A)} + \varphi_{(m-1,B)(n,A)} + \varphi_{(m-1,C)(n,A)}] + U\delta_{mn}\varphi_{(m,A)(n,A)}, \quad (2)$$

It is noted that there are totally nine eigen-equations in the 1D two-boson lattice model with three sublattices, see Ref. [39] for details. We note that the above eigen-equation of 1D two-boson system can be interpreted as describing a single bosonic particle hopping in a 2D lattice. The sites in the 2D lattice are labeled with four indices $(m,\alpha)(n,\beta)$, where $(m,\alpha)$ would be the position of the first particle, and $(n,\beta)$ corresponds to that of the second one. In this way, the diagonal lattice site $(m,\alpha)(m,\alpha)$ correspond to the two-boson state with double occupation. The interaction energy $U$ can be re-interpreted as an on-site potential in the diagonal lattice sites of the 2D model. In this case, there is an accurate correspondence between eigenequations of two correlated bosons in a 1D lattice and that of a single particle in the mapped 2D lattice.

Based on the similarity between circuit Laplacian and lattice Hamiltonian [40-58], electric circuits can serve as a very flexible platform to implement the mapped 2D lattice. The designed 2D circuit is plotted in Fig. 1c. Here, four circuit nodes connected by capacitors $C$ are used to form an effective lattice site in the lattice model. To simulate the real-valued hopping rate, four capacitors $C$ are used to directly link adjacent nodes without a cross, as shown in Fig. 1d. For realizing the hopping rate with a phase factor $\theta = \pm\pi/2$, four adjacent nodes are cross-connected by capacitors $C$, as shown in Fig. 1e. Each node is grounded by an inductor $L_g$, and the capacitors $C_U$ are used to ground nodes on the diagonal, as plotted in Figs. 1f and 1g. Through the appropriate setting of grounding and connecting, the circuit eigen-equation can be derived. For simply, the circuit eigen-equation related to Eq. (2) is displayed as:

$$\left(\frac{f_0^2}{f^2} - 10\right)V_{(m,A)(n,A)} = -[e^{i\frac{\pi}{2}}V_{(m,A)(n,B)} + V_{(m,A)(n,C)} + V_{(m,A)(n-1,C)} + V_{(m,A)(n-1,B)}e^{i\frac{\pi}{2}}$$
$$+V_{(m,B)(n,A)} + V_{(m,C)(n,A)} + V_{(m-1,B)(n,A)} + V_{(m-1,C)(n,A)}] + (C_U/C)\delta_{mn}V_{(m,A)(n,A)}, \quad (3)$$

where $f$ is the eigen-frequency ($f_0 = 1/2\pi\sqrt{CL_g}$) of the designed electric circuit and $V_{(m,\alpha)(n,\beta)}$ represents the voltage at the circuit node $(m,\alpha)(n,\beta)$. The details for the derivation of the circuit eigen-equations are provided in Ref. [39]. It is clearly shown that the eigen-equations of the designed circuit have the same form to that of the one-dimensional two-boson model, where the eigenenergy ($\varepsilon$) of two bosons is related to the eigen-frequency ($f$) of the designed circuit as $\varepsilon = f_0^2/f^2 - 10$. Other parameters are represented by $J=1$, $U=C_U/C$. In such an analogy, the probability amplitude of one boson at $(m,\alpha)$ and the other at $(n,\beta)$ for the 1D two-boson model is directly mapped to the voltage of the circuit node labeled by $(m,\alpha)(n,\beta)$. Specifically, the circuit nodes on the diagonal ($m = n$ and $\alpha = \beta$, dark purple nodes) and other sites ($m \neq n$ or $\alpha \neq$

$\beta$, light purple nodes) correspond to two bosons located at the same site and two distant ones, respectively, as shown in Fig. 1c. We note that it is difficult to simulate the two-boson model on a 1D circuit, where the tunable and strong interaction between two bosons are difficult to be emulated. Since the designed 2D circuit has a good correspondence with the 1D two-body Bose-Hubbard model, the designed 2D circuit can effectively simulate the behavior of two correlated bosons in the 1D lattice. In this case, we can map the low-dimensional few-particle quantum model to the high-dimensional electrical circuit network, where the number of quantum particles equals to the dimension of the circuit. However, a multi-particle model, such as a half-filled model with a large lattice length, would correspond to a very high-dimensional circuit. This is still very difficult to achieve in experiments

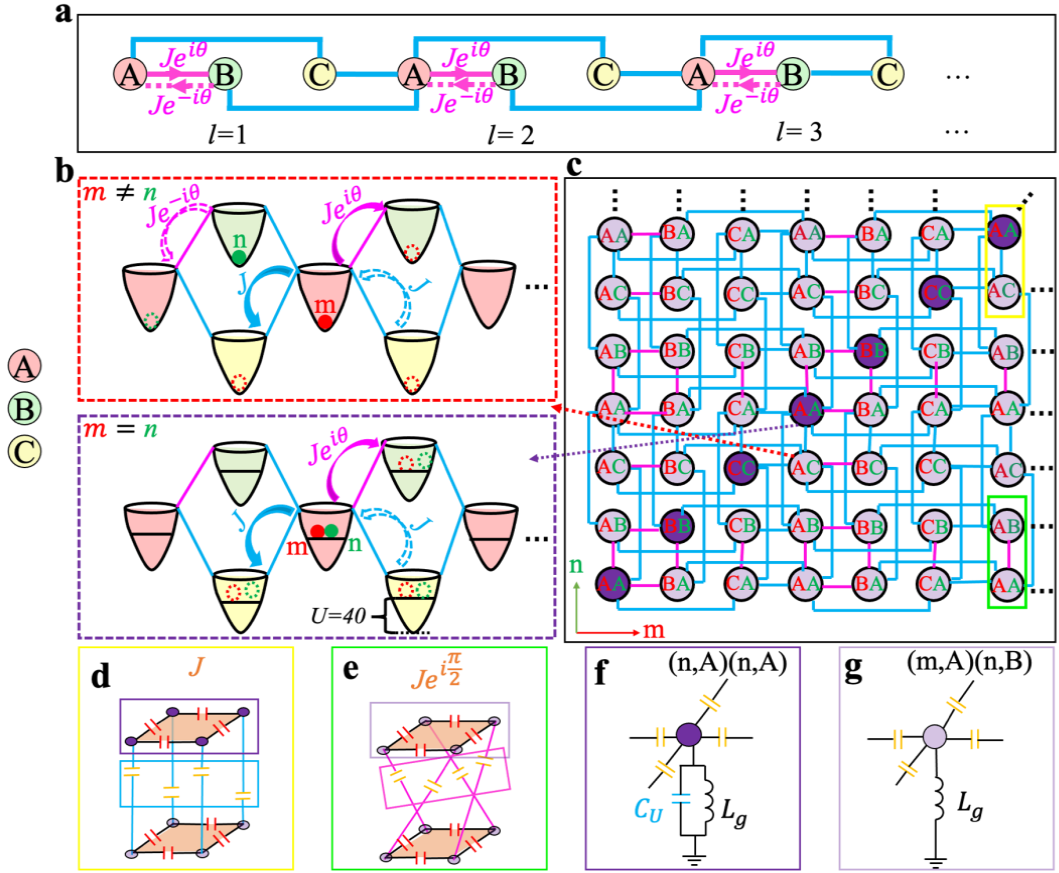

**FIG. 1. The electric circuit for simulating interaction-induced localizations and topology of two bosons. a.** Schematic diagram of a quasi-1D lattice with open boundary conditions. Each sublattice is marked with $(l, \alpha)$, where $\alpha = A, B, C$. A synthetic flux with $\theta = \pi$ is applied in each closed loop. **b.** Schematic diagram of quantum model of two correlated bosons. **c.** Schematic diagram of the designed 2D circuit lattice for simulating 1D two-boson model. **d.** and **e.** Illustrations of complex-valued and real-valued hopping rates in circuits between two sites enclosed by yellow and green blocks in Fig. 1c. **f.** and **g.** The detailed ground setting of circuit nodes enclosed by frames with consistent colors.

There are some interesting phenomena existing in the above model, where the two-boson flat-band localizations and topological edge states could be induced by strong interactions. To analyze the two-boson flat-band localization and topology, the eigen-spectrum of the designed circuit with $L = 55$ is calculated, as shown in Fig. 2a. The corresponding parameters are set as $C = 1nF, C_U = 40nF, L_g = 3.3uH$, respectively. The eigen-spectrum in the range of [0.71089MHz, 1.26313MHz] corresponds to the two-boson scattering state without strong interactions, where the significant dispersion exists due to the lacking of destructive inferences with a $\pi/2$ effective flux. Fig. 2d illustrates the spatial distribution of circuit eigenmodes related to the two-boson scattering states. We can see that the voltage amplitude is nearly disappeared on the circuit diagonal, corresponding to the state of two bosons located at distant sites. Fig. 2b displays the magnification of eigen-frequencies of two strongly interacted bosons at the same site. It is clearly shown that three flat bands at frequencies 0.39166MHz (green dots), 0.39142MHz (red dots) and 0.39079MHz (yellow dots) exist. Those flat bands all have a vanishing group velocity, meaning that two strongly interacted bosons possess the significant spatial localization. In addition, we plot the spatial distributions of the voltage amplitudes for two-boson modes at these three flat bands, as shown in Figs. 2e-2g. It is clearly shown that the voltage signals are all concentrated on the diagonal, corresponding to the state of two bosons located at the same site. It is worth noting that the coupling between strongly interacting two-boson states ($\varepsilon \sim U \gg J$) and two-boson scattering states ($\varepsilon \sim J$) can be ignored. Hence, the effective model of two bosons with strong interactions can be viewed as a 1D superlattice with three sublattices of $C_{(n,A)(n,A)}$, $C_{(n,B)(n,B)}$ and $C_{(n,C)(n,C)}$. The flat-band effect of two bosons with strong interactions is caused by the destructive interference in an effective two-boson AB cage. To further illustrate the mechanism of two-boson localizations, we draw the effective schematic diagrams for the destructive interference of two strongly-interacted bosons, as shown in Fig.2c. If we assume that the initial state of two bosons is located at the (n,A)(n,A) node, there are eight paths for such two-boson state to reach to the (n+1,A)(n+1,A) node of the neighbor unit. In this case, we can control the interference behavior of these eight paths by tuning the Peierls' phase factor of the single-particle hopping strength. It is found that a complete destructive interference can be formed with $\theta = \pi/2$ (an effective $\pi$-flux in each loop), which is the origin for the appearance of interaction-induced two-boson flat bands.

Another interesting feature of the above 1D two-boson lattice model is the existence of interaction-induced non-trivial topological edge states. Previous investigations haves shown that the $\pi$-flux in Aharonov-Bohm cages can induce the creation of square-root topological edge states [29]. It is shown in our work that the strongly interacting two-boson system with the single-particle hopping phase being $\theta = \pi/2$ is equivalent to an effective Aharonov-Bohm cage with a $\pi$ magnetic flux. In this case, the topological edge states of doublons can exist in our model even the single-particle counterpart is trivial. The midgap two-boson eigenmodes located in two-boson flat-band gaps (labeled by green and pink regions) correspond to the two-boson topological edge states. To further illustrate those two-boson topological edge states, we plot the corresponding spatial distributions of midgap eigenmodes, as presented in Figs. 2h and Figs. 2i. It is clearly shown that the voltage amplitudes are strongly concentrated on edges of the 2D circuit, manifesting the existence of two-boson topological edge states. In Ref. [39], we numerically calculate the eigen-spectra of two-boson model with different interaction strengths, and find that the two-boson flat bands and topological edge states are disappeared with weak interactions.

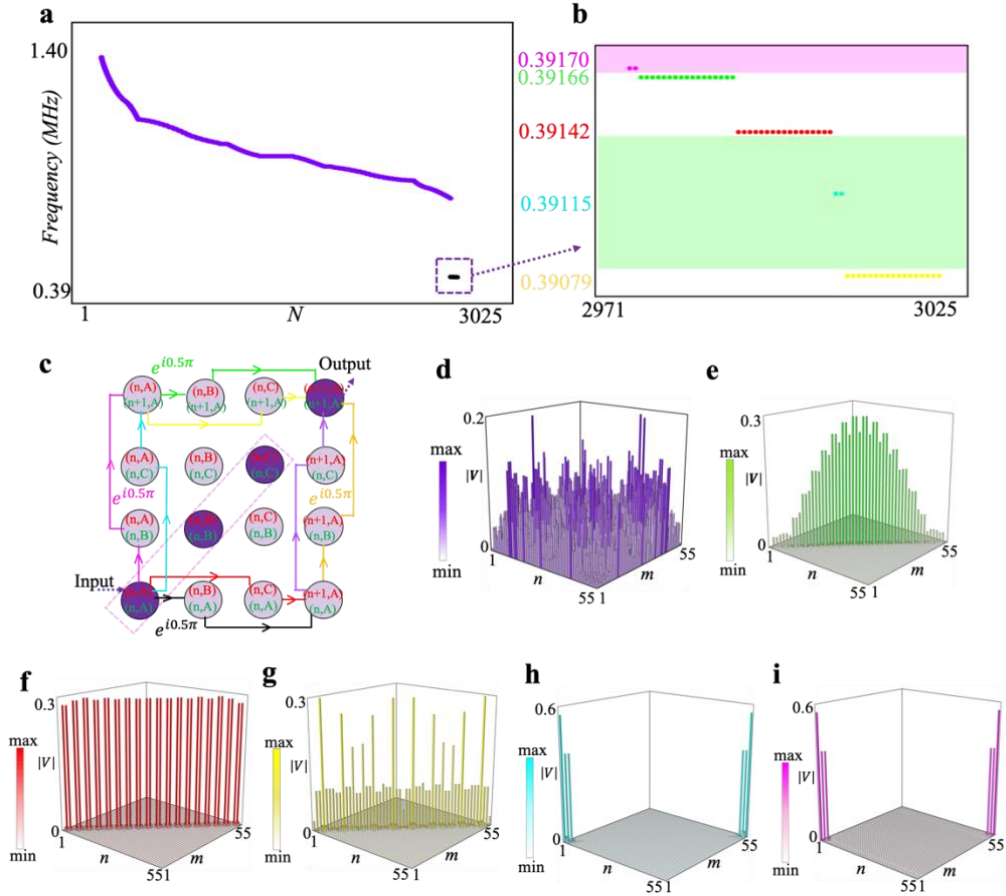

**FIG. 2. Numerical results of interaction-induced two-boson localizations and topological edge states in electric circuits. a.** Eigen-frequencies of the designed electric circuit with $m \times n = 55 \times 55$, and other parameters are set

as $C = 1nF, C_U = 40nF, L_g = 3.3uH$. **b.** Enlarged view of eigen-frequencies with interactions. **c.** Aharonov-Bohm cage. In the system with $\theta = \pi/2$, the time evolution of the state $|1, A, 1, A\rangle$ never allows the particle to leave the Aharonov-Bohm cage centered on that site (shaded in dark purple). **d.** The voltage distribution at $f$=0.897MHz. **e-i.** The voltage distribution diagram, the corresponding frequency is consistent with Fig 2b.

**Observing flat-band localizations and topological states induced by effective strong interactions by circuit networks.** In this part, we experimentally emulate the interaction-induced flat-band localizations and topology based on electric circuits. Here, the value of $C, C_U, L_g$ are taken as $C = 1nF, C_U = 40nF, L_g = 3.3uH$, respectively. Front and back sides for the photo of fabricated 2D circuit are shown in Fig. 3a and Fig. 3b. The circuit sample contains nine units. Four circuit nodes correspond to an equivalent lattice point, which are connected by capacitor $C$ (red). In this case, if four pair of adjacent circuit nodes are directly (cross) connected through the capacitors $C$ (orange), the hopping rate without (with) a phase-factor $e^{\pm i\pi/2}$ could be realized. The white dotted line in Fig. 3b marks the diagonal of the board, and the ground capacitance corresponding to the two-boson interaction is enclosed by the blue box. The ground inductance is marked with a pink box. Furthermore, we select the low-loss inductors, and different circuit elements are chosen to have low tolerances to avoid detuning of the circuit response (the fluctuations of circuit elements at and away the diagonal are below 1% and 5%).

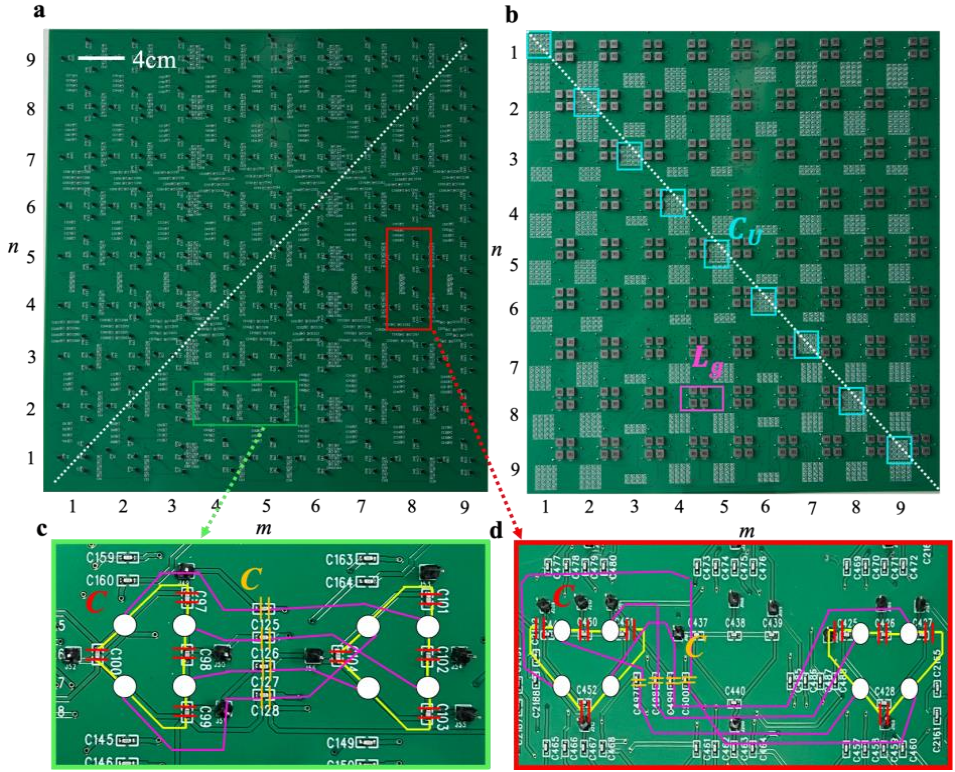

**FIG. 3 The fabricated circuit sample to simulate the interaction-induced localizations and topological edge

**states. a.** The photograph image of the front of the fabricated circuit simulator. **b.** The photograph image of the back of the fabricated circuit simulator. **c.** and **d.** Enlarged views of the green and red boxes in Fig. 3a.

It is well known that the impedance response of a circuit node is related to the local density of states of the corresponding quantum lattice model. In this case, in order to verify the phenomenon of two-boson flat-band effects, we measure the impedance spectrum of the (2, A)(2, A) bulk node at diagonal, as shown in Fig. 4a. We can see that two impedance peaks at 0.3907MHz (marked by the yellow dotted line) and 0.39145MHz (marked by the green dotted line) can be observed. The values of these two peaks are consistent to the eigenfrequencies of two-boson flat-band modes calculated in Fig. 2b, indicating the effective excitation of two-boson flat-band modes in 2D circuit. There is a little deviation between frequencies of impedance peaks in Fig. 4a and calculated eigenfrequencies in Fig. 2b. This is caused by the finite-size effect. In Ref. [39], we provide detailed simulation results of impedance spectra of circuits with different sizes. The associated simulation result (by LTSpice) is plotted in Fig. 4c with the blue line. We can see that simulated frequencies of two impedance peaks are matched to experimental results. The wide impedance peaks are due to the large lossy effect in experiments. And, due to the smaller magnitude of the impedance peak around 0.39145 MHz marked by the green line, it is much less sharp compared to simulations with low losses. Moreover, we also measure the impedance spectrum at the (2, B)(2, B) node, as displayed in Fig. 4b. We can see that three impedance peaks at 0.3907MHz (marked by the yellow dotted line), 0.3914MHz (marked by the red dotted line) and 0.39145MHz (marked by the green dotted line) are matched to eigenenergies of three flat-band doublon states in Fig. 2b. The orange line in Fig. 4c corresponds to the simulated impedance spectrum of (2, B)(2, B) node, where the frequencies of impedance peaks are the same to that of measurements. It is worth noting that the lack of central impedance peak of (2, A)(2, A) node is due to the zero-amplitude at A sublattices of the eigenmode for the central flat band. Then, we measure the impedance spectrum at (1, C)(2, B) bulk node away from the diagonal, which corresponds to the two-boson states with the single occupation, as displayed in Fig. 4d. The corresponding simulation result is shown in Fig. 4e. The widening of measured impedance peaks is caused by the lossy effect in the circuit sample. In addition, the deviation of frequencies and values of impedance peaks is resulting from disorders in the sample, which mainly come from the fluctuation of circuit elements and parasitic capacitances in PCBs. While, the measured impedance peaks are still located in the frequency range from 0.7 MHz to 1.3 MHz, that is matched to the eigen-energies of two-boson scattering states. In particular, a large number of impedance peaks indicate that the two-boson scattering states exist in a dispersive energy band. It is important to note that the weaker localization of the associated two-boson eigen-state is, the disorder-induced deviation of impedance spectra is more significant. Hence, we can still find a good consistence between measurements and simulations of strongly localized two-boson flat-band states, where the fluctuation of circuit elements at diagonal is limited below 1%.

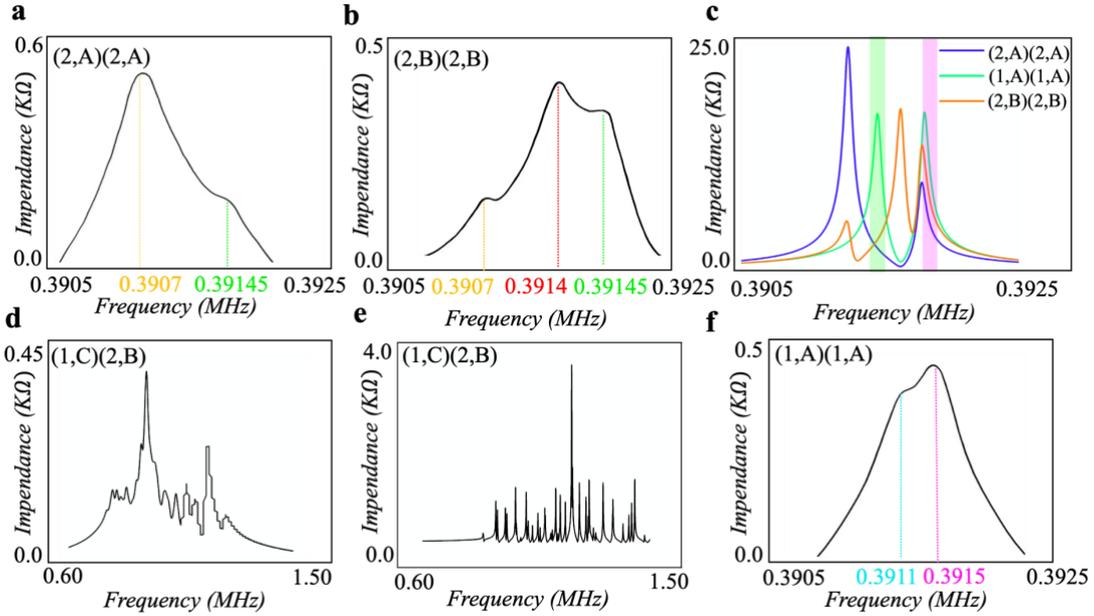

**FIG. 4. Experimental results of frequency-dependent impedance spectra on the simulation of interaction-induced two-boson flat bands and topological edge states. a, b, d, f.** The experimental impedance responses of (2, A)(2, A), (2, B)(2, B), (1, C)(2, B) and (1, A)(1, A) circuit nodes. **c. and e.** The simulated impedance responses of different bulk and edge nodes.

To further detect the two-boson topological edge states, we measure the impedance spectrum of the (1, A)(1, A) boundary node, as shown in Fig. 4f. It is shown that two impedance peaks at frequencies 0.3911MHz (blue dotted line) and 0.3915MHz (pink dotted line) can be observed, that are matched to the eigenfrequencies of two-boson topological edge states in Fig. 2b. The simulated impedance spectrum of (1, A)(1, A) boundary node is presented by the green line in Fig. 4c. A good consistence of measured and simulated frequencies of impedance peaks for the edge node is obtained. Similarly, the larger widths of measurements are resulting from significant losses in fabricated circuits. And, the unequal magnitude of two impedance peaks is induced by the weak disorder at the diagonal of the circuit and the dispersion effect of circuit elements. In Ref. [39], we provide detailed simulation results of impedance spectra with different strengths of losses, which are quantized by the effective series resistance of inductance. We find that the effective series resistance of inductance is about 100m$\Omega$ in our fabricated circuit sample. The above results clearly shown that the interaction-induced flat bands and topological edge states have been effectively simulated by designed 2D circuit networks.

Except for the unique impedance response, the existence of flat-band topology and localization can also exhibit exotic voltage dynamics. To further demonstrate these interesting phenomena, we perform time-domain measurements of our designed 2D circuit. Here, the circuit excitation is in the form of $[V_1 = V_0 e^{i\omega t}, V_2 = iV_0 e^{i\omega t}, V_3 = -V_0 e^{i\omega t}, V_4 = -iV_0 e^{i\omega t}]$. The angular frequency $\omega$ of the input voltage signal should be matched to the corresponding eigen-frequency of the designed circuit. Details of the experimental measurement are provided in Ref. [39].

Firstly, to study the localization phenomenon caused by the two-boson flat band effect, we input the voltage signal at the (2,A)(2,A) node with $\omega = 2.4554MHz$ (matched to the two-boson flat-band bulk mode), and measure the voltage dynamics from 0 to 5$ms$, as shown in Fig. 5a. We can see that the injected voltage signal is strongly localized on several circuit nodes near the excitation point. To further see the distribution of the voltage signal, we plot the voltage distribution at 3$ms$, as shown in Fig. 5b. It is clearly seen that the voltage signal is concentrated at five circuit nodes as (1,B)(1,B), (1,C)(1,C), (2,A)(2,A), (2,B)(2,B) and (2,C)(2,C), being consist to the localization pattern of single-particle AB cage. To further illustrate that the voltage localization is caused by the flat-band effect in the strongly correlated two-boson system, we also measure the voltage evolution by exciting the (1,C)(2,B) circuit node at $\omega = 5.5104MHz$ (matched to the two-boson scattering states), as shown in Fig. 5c. Fig. 5d displays the corresponding spatial profile of voltage signal at 3$ms$. We can see that the voltage signal quickly diffuses into the whole sample, indicating that the spatial localization is absence for two-boson scattering states.

Then, in order to measure the voltage dynamics related to two-boson topological edge states, we excite the circuit at the (1,A)(1,A) node with $\omega = 2.4611MHz$ (matched to the two-boson topological edge mode). Due to the existence of topological edge states, the input voltage signal is always localized on the corner of 2D circuit, and does not spread over the entire system, as shown in Fig. 5e. Similarly, we also plot the distribution of the voltage signal at 3$ms$, as shown in Fig. 5f. We can clearly see that the voltage signal is concentrated on the boundary unit at (1,A)(1,A),

(1,B)(1,B) and (1,C)(1,C) nodes. Those results clearly demonstrate the existence of topological edge state. Finally, it is worthy to note that the above experimental results are also consistent with simulation results, provided in Ref. [39].

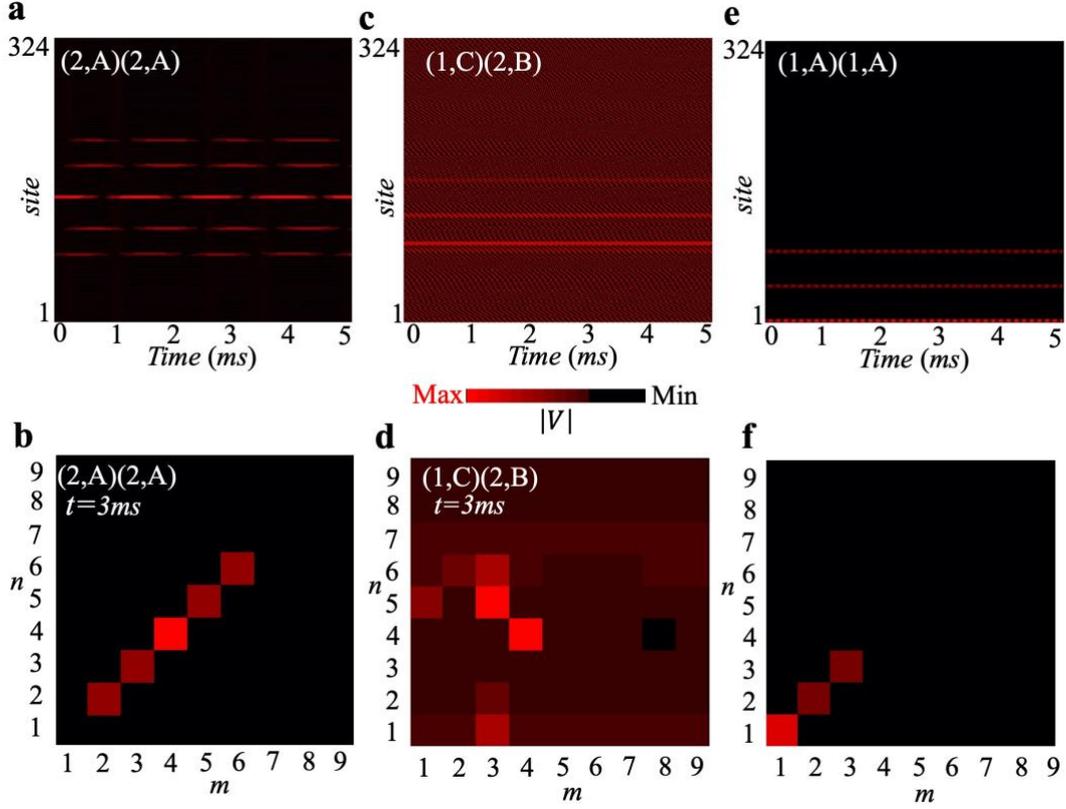

**FIG. 5. Experimental results of voltage dynamics. a, c and e.** The experimental results of the time dynamics of (2,A)(2,A), (1,C)(2,B) and (1,A)(1,A) nodes in the fabricated circuit, respectively. **b, d and f.** The distribution of the voltage amplitude at t=3*ms* related to results in Figs. 5a-5c, respectively.

**Conclusion.** In conclusion, we have experimentally emulated the flat-band localizations and topological edge states induced by effective strong interactions in circuit networks. In particular, the eigenstates of two correlated bosons in the one-dimensional AB cage have been mapped to eigenmodes of designed 2D circuit lattices. By tuning the effective interaction strength through circuits' groundings, the two-boson flat-bands and topological edge states have been detected by measuring frequency-dependent impedance responses and voltage dynamics. With the flexibility that the connection and grounding of circuit nodes are allowed in any desired way free from constraints of locality and dimensionality, circuit networks could provide a flexible platform to further investigate few-body topological flat bands beyond two bosons. In addition, the few-body topological flat-band systems with non-Hermitian and non-linear properties can also be simulated by suitably designed circuit networks. Our proposal provides a flexible platform to investigate and visualize many interesting phenomena related to interaction-induced flat-band topology, and also gives a way to manipulate the electronic signals with exotic behaviors.

**ACKNOWLEDGMENTS**.

This work was supported by the National Key R & D Program of China under Grant No. 2022YFA1404900 and the National Natural Science Foundation of China under Grant No.12104041.